\documentstyle[emulateapj,psfig]{article}
\begin{document}

%
\newcommand{\ltsim}{\lower.5ex\hbox{$\; \buildrel < \over \sim \;$}}
\newcommand{\gtsim}{\lower.5ex\hbox{$\; \buildrel > \over \sim \;$}}
\newcommand{\order}[1]{\mbox{$\cal{O}$ ({#1})}}
\newcommand{\etal}{\mbox{\it et~al.}}
\newcommand{\suph}{\mbox{$^{\rm h}$}}
\newcommand{\supm}{\mbox{$^{\rm m}$}}
\newcommand{\sups}{\mbox{$^{\rm s}$}}
\newcommand{\degree}    {\mbox{$^\circ$}}

\def\arcmin{\hbox{$^\prime$}}
\def\arcsec{\hbox{$^{\prime\prime}$}}
\def\degs       {$^\circ$}
\def\grs        {$\gamma$-rays }
\def\gr         {$\gamma$-ray }
\def\pers       {$\rm s^{-1}$}
\def\ergps      {$\rm erg$ ~\pers }
\def\etal     {$et~al.$}
\def\apr{$\sim$}
\def\asca {$ASCA\,$}
\def\fluxunit {erg cm$^{-2}$s$^{-1}$kev$^{-1}$}
\def\etal {{\em et al.}}
\def\spose#1{\hbox to 0pt{#1\hss}}
\newcommand\lsim{\mathrel{\spose{\lower 3pt\hbox{$\mathchar"218$}}
     \raise 2.0pt\hbox{$\mathchar"13C$}}}
\newcommand\gsim{\mathrel{\spose{\lower 3pt\hbox{$\mathchar"218$}}
     \raise 2.0pt\hbox{$\mathchar"13E$}}}

\newcommand{\Msun}      {\mbox{$\,M_{\mathord\odot}$}}
\newcommand{\Lsun}      {\mbox{$\,L_{\mathord\odot}$}}
\newcommand{\Rsun}      {\mbox{$\,R_{\mathord\odot}$}}

\newcommand{\eg}{\mbox{\it e.g.~}}
\newcommand{\xte}{\mbox{XTE\,J1118+480}}

\title{USA and {\em RXTE} Observations of a Variable Low-Frequency QPO
in \xte}

\author{K. S. Wood\altaffilmark{1}, P. S. Ray\altaffilmark{2},
R. M. Bandyopadhyay\altaffilmark{3}, M. T. Wolff, \\
\vspace{0.8mm}
G. Fritz, P. Hertz\altaffilmark{4}, M. P. Kowalski, M. N. Lovellette,  
D. Yentis}
\vspace{0.5mm}
\affil{E. O. Hulburt Center for Space Research, 
	Naval Research Laboratory,
	Washington, DC 20375}
\vspace{0.7mm}
\author{E. Bloom, B. Giebels, G. Godfrey, K. Reilly, P. Saz Parkinson,
G. Shabad} 
\affil{Stanford Linear Accelerator Center, Stanford University, 
	Stanford, CA 94309}
\and
\author{J. Scargle}
\affil{NASA/Ames Research Center, Moffett Field, CA 94035}
\altaffiltext{1}{Kent.Wood@nrl.navy.mil}
\altaffiltext{2}{Paul.Ray@nrl.navy.mil}
\altaffiltext{3}{NRL/NRC Research Associate}
\altaffiltext{4}{Current address: NASA Headquarters, Washington, DC}
\begin{abstract}
The USA experiment on {\em ARGOS} and {\em RXTE} have exensively
observed the X-ray transient \xte\ during its recent outburst in 2000
April--June.  We present detailed monitoring of the evolution of a low
frequency QPO which drifts from 0.07 Hz to 0.15 Hz during the
outburst.  We examine possible correlations of the QPO frequency with
the flux and spectral characteristics of the source, and compare this
QPO to low frequency QPOs observed in other black hole candidates.
\end{abstract}


\section{Background on \xte}

The transient source \xte\ (\cite{rmss00}) in Ursa Major
appeared as an X-ray source in the first few days of 2000 and has
continued to be detectable through the subsequent six months.  Its X-ray
activity to date has consisted of two distinct outbursts, the first of
which rose to maximum rapidly and declined in 2000 January.  The rise
to the second maximum followed about 25 days after the decline from
the first and it was during this second maximum that the All Sky
Monitor (ASM) on the {\em Rossi X-ray Timing Explorer} ({\em RXTE})
discovered the source.  The discovery announcement noted the earlier
maximum based on pre-discovery observations.

\xte\ is tentatively classified as a black hole transient based on the
detection of low-frequency quasi-periodic oscillations (QPOs), lack of
high-frequency noise (\cite{rsb00}), and power-law photon spectrum
extending to at least 120 keV (\cite{wm00}).  However, no dynamical mass
measurement has yet been made for the compact object, preventing
definitive identification as a black hole.  The X-ray flux is low, not
having exceeded 75 mCrab at any time to date.  Tentative distance
estimates to the putative optical counterpart based on companion
spectral types in other X-ray transients (\cite{ukm+00}) imply an
unusually low X-ray luminosity ($L_X \sim 10^{34}-10^{35}$ erg/s).
Its Galactic latitude of 62\degree\ is extremely high for this source
class, and the nominal distance would place it in the Galactic halo
(\cite{ukm+00}).  The second outburst has prompted extensive
multi-wavelength monitoring, including observations in radio, optical,
EUV, soft X-rays and hard X-rays.  A continuous spectrum is inferred
from optical to hard X-rays with the photon spectral index determined
in X-rays as $1.8 \pm 0.1$ (\cite{hmh+00}).

X-ray variability of the source is characterized in the time domain by
repeated flaring on timescales of a few seconds.  The power spectrum
shows a variable, low-frequency QPO near 0.1~Hz, first reported from
observations with the Proportional Counter Array (PCA) on {\em RXTE}
(\cite{rsb00}).  An outstanding feature of the source is that the QPO
is seen in X-rays, EUV and optical (\cite{hsp+00}).  The optical
source shows a photometric period, presumably orbital, of 4.1 hours
(\cite{ukm+00}). During the outburst the source has brightened from $V
\gtsim 18.8$ in quiescence to $V \approx 13$ (\cite{ukm+00}).

This {\em Letter} reports X-ray observations made with the
Unconventional Stellar Aspect (USA) experiment on the US Air Force
{\em Advanced Research and Global Observation Satellite} 
({\em ARGOS}) and the PCA on {\em RXTE}.  The $\sim 0.1$Hz QPO is 
detected in most USA and {\em RXTE} observations, providing a picture
of the evolution of this QPO over more than 60 days with unprecedented
coverage.  The USA observing sequence was initiated as soon as
possible after the first report (\cite{rmss00}) of the transient and
has subsequently been maintained through the remainder of the second
maximum of the outburst. The high revisit frequency of USA
observations is meant to provide, among other things, a framework for
linking up and correlating multi-wavelength observations.

\section{Description of the USA experiment}

USA is an X-ray timing experiment built jointly by the Naval Research
Laboratory and the Stanford Linear Accelerator Center for the dual
purposes of conducting studies of variability in X-ray sources and
exploring applications of X-ray sensor technology (see Ray \etal\
1999\nocite{rwf+99} for a more detailed description).  USA was
launched on 1999 February 23 on {\em ARGOS} into a nearly circular
830~km orbit at 98.8\degree\ inclination. It is a reflight of two
proportional counter X-ray detectors flown previously on the NASA {\em
Spartan-1} mission (\cite{kcs+93}), after which they were recovered
and refurbished.  The primary observing targets are bright Galactic
X-ray binaries, with a goal of obtaining large exposures on a modest
number of sources.

The detector consists of two multiwire constant-flow proportional
counters equipped with a 5.0~$\mu$m Mylar window and an additional
1.8~$\mu$m thick aluminized Mylar heat shield. The detector gas is a
mixture of 90\%\ argon and 10\%\ methane (P-10) at a pressure of 16.1
psia (at room temperature).  The detectors are sensitive in the range
1--15 keV with an effective area of about 1000 cm$^2$ per detector at
3 keV.  The collimators serve to support the window as well as to
define the field of view, which is approximately 1.2\degree\ FWHM
circular.  The Crab Nebula gives about 4000 cts/s in one detector at
the center of the field of view.  The {\em ARGOS} spacecraft on which
USA is mounted is three-axis-stabilized and nadir-pointed.
Consequently, the X-ray detectors are mounted on a 2-axis gimballed
platform to permit inertial pointing at celestial objects.  The pitch
and yaw drive capability is $\sim 3.5$\degree /min (track) and $\sim
20$ \degree/min (slew) and the common pitch/yaw pivot design allows
180\degree\ travel in each axis.

USA has five telemetry modes, four event and one spectral.  Events are
time-tagged to an onboard GPS receiver which provides an absolute time
reference.  Events are recorded with either 32 $\mu$s time resolution
and 16 pulse height channels or 2 $\mu$s resolution with 8 channels.
Event modes can be used up to count rates of $\sim$1000 cts/s at 40
kbps or $\sim$6000 cts/s at 128 kbps before events are discarded.
Spectral mode records a 48 channel spectrum every 10 ms.

The USA instrument has performed well since activation began on 1999
April 30, but the mission has not been without difficulties.
Approximately two weeks after launch the detector heat shields
suffered from degradation which has imposed additional constraints on
USA pointing with respect to the Sun. On 1999 June 8, Detector 2
developed a rapid gas leak, possibly caused by a micrometeor impact,
and exhausted its gas supply leaving only Detector 1 to complete the
mission, halving the effective area.

\section{Observations and Data Analysis}

USA has obtained a total of 425 ks of data on \xte\ between 2000 April
10 and June 13.  These observations are continuing through the duration
of the outburst.  Using the 32 $\mu$s event mode, the source was
observed between 5 and 11 times per day, providing the most consistent
sampling of its X-ray behavior in existence.  We have also utilized
the 24 public target-of-opportunity (TOO) observations of \xte\ made
by {\em RXTE} between 2000 April 13 and June 11.  Analysis of the USA
and {\em RXTE} data, including energy selection, binning, construction
of the power spectra, and energy spectral fitting, was performed using
FTOOLS v5.0.1.  Typically the observed QPO has a fractional RMS
amplitude of $\sim 5$\% and a FWHM of $\sim$0.01~Hz.  Additionally, in
some cases there are suggestions of multiple peaks in the power
spectra.  Detailed results of our QPO analysis are summarized in Table~1.

\subsection{USA}

To search for QPOs in the USA data, we created 30 groups of
observations, each of which contain 4--8 individual observations from
within a continuous interval of 8--20 hours.  As a result of the nearly polar
orbit of USA, the usable time from each individual observation varies
in length between 300 and 1100 seconds; hence the total on-source
integration time contained in each group ranges from 1.8 to 4.9 ks.
To construct the power spectrum for each group of USA observations, we
selected channels 1--8, corresponding to an energy range of 1--10 keV,
and binned the data by either 0.1 or 0.125 seconds.  Power spectra of
length 4096 were computed and averaged for each observation, the
expected Poisson level subtracted, and normalized to fractional
RMS$^2$/Hz \nocite{nvw+99} (see Nowak \etal\ 1999 and references
therein for a detailed description of power spectral analysis of X-ray
data). The QPO parameters were determined by fitting the resultant
power spectra between 0.03 and 0.8 Hz to a power law plus a Gaussian
QPO feature.  The centroid and FWHM of the feature were determined
from the fit parameters and the fractional RMS attributed to the QPO
was determined from the integral of the Gaussian profile.  USA average
fluxes for each observation are determined by subtracting a background
model and correcting for obscuration of the detector by the instrument
support structure.  The USA spectral calibration is still being
determined and spectral results from USA will be presented in a future
paper.

\subsection{RXTE}

The power spectra from the {\em RXTE} observations were generated from
the {\tt Standard1} data which bin all of the good counts in each
detector at 0.125 s (\cite{jsg+96}).  These time series were summed
for all active PCUs and the power spectra were computed and fitted as
described above for the USA data.  Spectral fits were also made using
the PCA data, employing the background models and response matrices
provided by the {\em RXTE} team.  We fit a pure power law model
between 2.5 and 20 keV to each of the RXTE observations listed in
Table 1.  This provided a reasonably good fit to the data, but the
residuals typically show excesses below 4 keV and near 6 keV.  Fitting
a power law between 8 and 25 keV to avoid these features, we find
that the photon spectral index is consistent with a constant value of
1.73$\pm$0.01.

\section{Discussion}

\subsection{Source Models}

Examining the options for the nature of this unusual X-ray transient,
we note that its emission characteristics and timescales require a
compact stellar object, but the hard X-ray spectrum excludes a white
dwarf. A neutron star model has not been supported by detection of
signatures such as pulsations, Type I X-ray bursts, or high frequency
QPO activity.  Therefore the observed X-ray characteristics of \xte\
seem to be best explained by identification of this source as a binary
black hole transient in the canonical low hard state.  However, this
source is far from a typical example of that class.  The luminosity is
relatively low, and it is also unusual for a black hole transient to make
its first appearance in the low hard state.  (In this regard Uemura
\etal\ (2000)\nocite{ukm+00} have suggested that the source is viewed
at high inclination, but there are no observed eclipses to support
this suggestion.)  The possible Galactic halo location of \xte\ and the
X-ray/EUV/optical QPO are unique, although previous transients of this
type have shown X-ray and optical correlated variability.  The ratio
of the X-ray to optical flux is also extremely low, leading to
speculation that this outburst of \xte\ may be a ``mini-outburst'' of
the type seen in the black hole candidate system GRO J0422+32
(\cite{hmh+00}).  We also note that short-timescale X-ray and optical
variability similar to that seen in \xte\ has been observed in the
hard state of the black hole candidate GX~339$-$4 (\cite{mot+83}).
Merloni {\etal}\ (2000)\nocite{mmf00} suggest that both the optical
and hard X-ray variability in \xte\ are due to magnetic flares in a
Comptonizing corona and predict X-ray/optical correlations which can
be tested with ongoing multi-wavelength observations. Alternatively,
Titarchuk and Osherovich (2000)\nocite{to00} argue that the low
frequency QPO is due to a global oscillation mode in the accretion
disk and show that these frequencies carry information about the
system as a whole such as the size of the disk and the mass of the
central object.

\subsection{QPO Evolution}

Figure 1 shows the evolution of the source flux as monitored by the
ASM on {\em RXTE} (\cite{lev+96}) and the QPO centroid as measured by
USA and {\em RXTE}. The QPO moves upward in frequency over the 62 days
of observations, apparently monotonically, but with significant
changes in rate.  During the same period, the 2--12 keV X-ray flux
slowly rises, then begins to decrease.  Figure~2 shows the QPO
centroid frequency versus measured USA (1--10 keV) fluxes.  Clearly, the
QPO frequency fails to track the source intensity according to a
simple proportionality.

This effect could be correlated with change in the source energy
spectrum.  However, the power-law component between 8 and 25 keV seen
with the PCA stays essentially constant while the QPO frequency
changes by a factor of two.  It is possible that a soft component,
which peaks below the PCA energy range, tracks the state of the
accretion disk and thus the QPO frequency.  The variable flux in the
power-law component would then be due to inverse-Compton scattered
seed photons from the variable disk component.

Low-frequency QPOs have been seen in several black hole binaries in the
low hard state, including Cyg X-1, GRO J0422+32, and GX339$-$4.  Unlike
the QPO observed in \xte\ , the $\sim$0.3~Hz QPO observed in the low
hard state of GX339$-$4 appears to be relatively stable, but like \xte\
no simple correlation between the QPO frequency and source flux was
found (\cite{nwd+99}).  Variable low-frequency QPOs have been observed
during the high and intermediate states of the black hole transient
XTE~J1550$-$564.  In this source, the QPO frequency generally increases
as the disk flux increases (\cite{som+99}).  However, we note that the
time evolution of the variable QPO in XTE~J1550$-$564 is more erratic
than in \xte\, with the QPO frequency increasing and decreasing
multiple times within a similar interval ($\sim$60 days).

\section{Summary}

The QPO in \xte\ shows an upward drift from 0.07 Hz to 0.15 Hz over a
62 day time interval while the source intensity slowly rises and then
decreases.  The effect is clear but the explanation seems to call for
changes in the disk state that are not completely specified by
measurement of the PCA X-ray spectrum, possibly by a variable
soft disk component which is not well determined by these observations.
We are further investigating the variability by studying flaring in
the time domain.  The flare events, which have a large signal-to-noise
ratio and occur on the same timescale as the QPO, may shed light on
the physical processes behind the QPO.  Continued X-ray observations will
be important for covering a larger dynamic range in the flux and
continuing to track the QPO centroid.

\acknowledgements

This paper is dedicated to the memory of Terry E. Crandall who
contributed many ingenious and creative solutions to the USA
Experiment software.

We would like to acknowledge Ganwise Fewtrell for his contributions to
the USA data analysis software.  Basic research in X-ray Astronomy at
the Naval Research Laboratory is supported by the Office of Naval
Research.  Work at SLAC was supported by Department of Energy contract
DE-AC03-76SF00515.  This work was performed while RMB held a National
Research Council Research Associateship Award at NRL.  This paper made
use of quick-look results provided by the ASM/{\em RXTE} team (see
{\tt http://xte.mit.edu}).


\begin{table*}
\caption{Parameters of low-frequency QPO in XTE J1118+480}
\begin{center}
\scriptsize
\begin{tabular}{rrccccrl}\hline
UT Start     & UT End	& T$_{obs}$ & Frequency 	& FWHM   	    & rms	    &Count rate\tablenotemark{a}& Instrument\\ 
       	     &		& (ks)    &  (Hz)     	& (Hz)   	    &($\%$)	    &(counts s$^{-1}$)&\\\hline
10 Apr 18:27 & 11 Apr 06:31 & 4.1  & 0.0689$\pm$0.0011	& 0.0115$\pm$0.0022 & 6.7$\pm$1.6 & 107.5 & USA \\
11 Apr 18:10 & 12 Apr 06:14 & 4.9  & 0.0716$\pm$0.0013	& 0.0138$\pm$0.0025 & 6.2$\pm$1.5 & 100.8 & USA \\
12 Apr 19:34 & 13 Apr 05:56 & 4.2  & 0.0724$\pm$0.0007	& 0.0064$\pm$0.0012 & 5.9$\pm$1.8 & 102.5 & USA \\
13 Apr 09:44 & 13 Apr 14:02 & 11.2 & 0.0769$\pm$0.0006  & 0.0109$\pm$0.0013 & 6.9$\pm$1.1 & 410.4 & XTE(3) \\
13 Apr 14:25 & 13 Apr 15:40 & 3.1  & 0.0774$\pm$0.0012  & 0.0098$\pm$0.0022 & 6.5$\pm$2.1 & 406.1 & XTE(3) \\
13 Apr 20:59 & 14 Apr 07:22 & 3.0  & 0.0790$\pm$0.0008	& 0.0058$\pm$0.0016 & 5.5$\pm$2.1 & 103.1 & USA \\
14 Apr 14:02 & 15 Apr 00:20 & 4.4  & 0.0818$\pm$0.0018	& 0.0094$\pm$0.0032 & 4.4$\pm$1.7 & 103.0 & USA \\
15 Apr 07:59 & 15 Apr 08:20 & 1.1  & 0.0818$\pm$0.0013  & 0.0125$\pm$0.0026 & 9.6$\pm$3.7 & 689.8 & XTE(5) \\
17 Apr 05:10 & 17 Apr 07:12 & 4.1  & 0.0838$\pm$0.0026  & 0.0318$\pm$0.0058 & 8.2$\pm$1.8 & 419.5 & XTE(4) \\
18 Apr 17:50 & 19 Apr 05:52 & 5.0  & 0.0884$\pm$0.0008	& 0.0106$\pm$0.0018 & 5.6$\pm$1.4 & 97.4  & USA \\
18 Apr 21:47 & 18 Apr 23:02 & 1.8  & 0.0861$\pm$0.0017  & 0.0146$\pm$0.0037 & 7.7$\pm$2.7 & 455.8 & XTE(3) \\
20 Apr 14:01 & 21 Apr 00:18 & 4.5  & 0.0827$\pm$0.0014	& 0.0158$\pm$0.0031 & 5.7$\pm$1.4 & 96.4  & USA \\
21 Apr 04:17 & 21 Apr 04:34 & 0.9  & 0.0873$\pm$0.0010  & 0.0060$\pm$0.0017 & 7.8$\pm$3.8 & 700.4 & XTE(5) \\
21 Apr 20:21 & 22 Apr 06:44 & 3.9  & 0.0884$\pm$0.0008	& 0.0062$\pm$0.0013 & 5.1$\pm$1.7 & 93.3  & USA \\
23 Apr 19:46 & 24 Apr 09:46 & 2.4  & 0.0885$\pm$0.0014	& 0.0095$\pm$0.0026 & 4.1$\pm$1.7 & 99.7  & USA \\
25 Apr 14:12 & 26 Apr 00:32 & 3.5  & 0.0980$\pm$0.0008	& 0.0089$\pm$0.0008 & 5.3$\pm$1.2 & 92.5  & USA \\
26 Apr 13:59 & 26 Apr 22:37 & 3.2  & 0.0930$\pm$0.0012	& 0.0149$\pm$0.0024 & 7.3$\pm$1.5 & 90.9  & USA \\
27 Apr 13:35 & 28 Apr 06:41 & 4.3  & 0.0956$\pm$0.0017	& 0.0225$\pm$0.0034 & 7.9$\pm$1.6 & 89.0  & USA \\
29 Apr 16:22 & 30 Apr 07:50 & 3.7  & 0.0955$\pm$0.0010	& 0.0089$\pm$0.0018 & 4.9$\pm$1.3 & 94.6  & USA \\
30 Apr 17:45 & 1 May 07:32  & 4.0  & 0.1029$\pm$0.0008	& 0.0089$\pm$0.0008 & 5.5$\pm$1.6 & 88.4  & USA \\
1 May 11:25  & 1 May 11:59  & 1.8  & 0.0994$\pm$0.0009  & 0.0097$\pm$0.0020 & 7.9$\pm$2.6 & 446.5 & XTE(3) \\
1 May 15:47  & 1 May 22:51  & 3.3  & 0.0979$\pm$0.0010	& 0.0108$\pm$0.0020 & 5.8$\pm$1.5 & 84.3  & USA \\
2 May 13:49  & 2 May 22:34  & 4.2  & 0.1010$\pm$0.0008	& 0.0080$\pm$0.0015 & 5.2$\pm$1.2 & 83.7  & USA \\
3 May 18:35  & 4 May 08:22  & 4.2  & 0.0990$\pm$0.0007	& 0.0080$\pm$0.0013 & 5.5$\pm$1.5 & 86.8  & USA \\
5 May 16:19  & 6 May 04:19  & 3.0  & 0.1068$\pm$0.0018	& 0.0213$\pm$0.0035 & 6.8$\pm$1.4 & 80.3  & USA \\
6 May 12:47  & 6 May 19:46  & 3.4  & 0.1106$\pm$0.0020	& 0.0122$\pm$0.0035 & 4.3$\pm$1.7 & 79.3  & USA \\
7 May 14:10  & 8 May 03:44  & 4.1  & 0.1135$\pm$0.0014	& 0.0140$\pm$0.0030 & 5.6$\pm$1.5 & 77.3  & USA \\
9 May 05:04  & 9 May 20:35  & 2.4  & 0.1133$\pm$0.0013	& 0.0075$\pm$0.0024 & 4.8$\pm$2.0 & 77.2  & USA \\
11 May 15:32 & 11 May 16:15 & 2.3  & 0.1127$\pm$0.0026  & 0.0241$\pm$0.0057 & 7.1$\pm$2.2 & 525.6 & XTE(4) \\
11 May 17:08 & 11 May 18:04 & 2.8  & 0.1122$\pm$0.0010  & 0.0108$\pm$0.0019 & 6.4$\pm$1.5 & 529.2 & XTE(4) \\
11 May 18:46 & 11 May 22:26 & 7.7  & 0.1184$\pm$0.0012  & 0.0187$\pm$0.0033 & 5.8$\pm$1.4 & 456.1 & XTE(4) \\
16 May 19:54 & 17 May 04:30 & 2.5  & 0.1187$\pm$0.0011	& 0.0136$\pm$0.0029 & 6.6$\pm$1.8 & 72.7  & USA \\
17 May 23:00 & 18 May 07:41 & 1.9  & 0.1140$\pm$0.0019  & 0.0109$\pm$0.0034 & 5.3$\pm$2.3 & 85.3  & USA \\
20 May 05:15 & 20 May 20:45 & 3.4  & 0.1173$\pm$0.0012	& 0.0075$\pm$0.0023 & 3.7$\pm$1.5 & 72.9  & USA \\
23 May 00:08 & 23 May 00:36 & 1.6  & 0.1188$\pm$0.0024  & 0.0148$\pm$0.0059 & 5.2$\pm$2.7 & 379.5 & XTE(3) \\
25 May 17:17 & 26 May 00:18 & 3.2  & 0.1272$\pm$0.0006	& 0.0078$\pm$0.0013 & 4.6$\pm$1.2 & 75.9  & USA \\
26 May 15:18 & 27 May 06:44 & 2.5  & 0.1240$\pm$0.0014  & 0.0183$\pm$0.0026 & 7.4$\pm$1.4 & 77.5  & USA \\
27 May 08:19 & 27 May 08:51 & 1.7  & 0.1171$\pm$0.0055  & 0.0526$\pm$0.0128 & 9.5$\pm$2.9 & 509.0 & XTE(4) \\
27 May 16:42 & 28 May 09:56 & 1.8  & 0.1273$\pm$0.0036  & 0.0247$\pm$0.0079 & 4.7$\pm$1.9 & 85.1  & USA \\
1 Jun 03:27  & 1 Jun 23:57  & 2.6  & 0.1275$\pm$0.0013	& 0.0130$\pm$0.0025 & 5.8$\pm$1.5 & 79.7  & USA \\
3 Jun 13:01  & 4 Jun 02:40  & 3.0  & 0.1247$\pm$0.0022	& 0.0278$\pm$0.0047 & 6.8$\pm$1.5 & 80.2  & USA \\ 
4 Jun 12:44  & 5 Jun 05:47  & 2.2  & 0.1341$\pm$0.0012  & 0.0122$\pm$0.0027 & 4.9$\pm$1.6 & 80.8  & USA \\
11 Jun 01:46 & 11 Jun 02:35 & 1.8  & 0.1591$\pm$0.0027  & 0.0244$\pm$0.0053 & 6.7$\pm$2.1 & 376.0 & XTE(3) \\\hline
\end{tabular}
\end{center}
\tablenotetext{a}{Average raw count rate, not background subtracted or
otherwise corrected, as used in the QPO analysis
(XTE points are Standard1 count rates for the number of detectors
indicated in parentheses).}
\end{table*}

\begin{figure*}
\centerline{\psfig{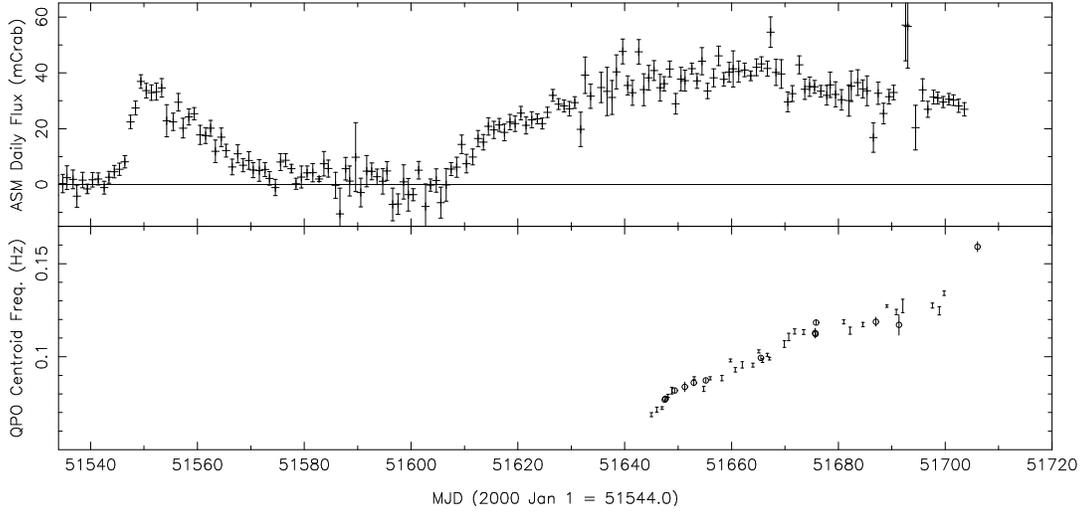}}
\caption{ASM daily average fluxes (top panel) and QPO centroid frequencies.
In the lower panel, capped error bars are QPO measured with USA, while open circles
represent QPO found in the {\em RXTE} PCA data.}
\end{figure*}

\begin{figure*}
\centerline{\psfig{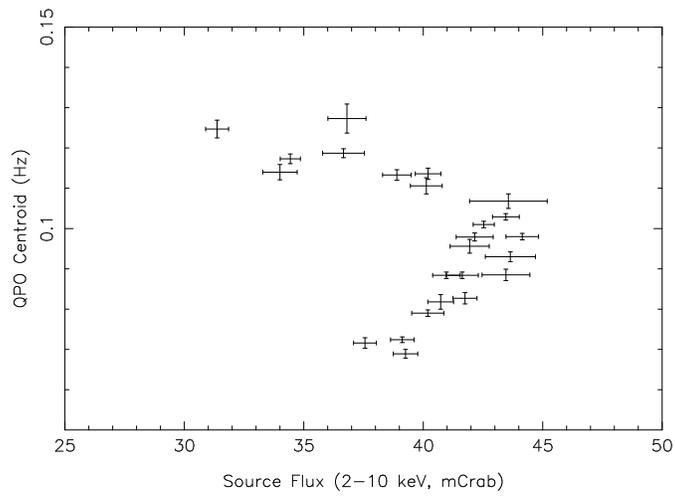}}
\caption{Measured QPO centroid frequency vs. source flux for each USA
observation of \xte.  Source fluxes are in mCrab, having been
background subtracted and corrected for obscuration of the detector by
the support structure.}
\end{figure*}

\end{document}